   \documentclass[preprint,authoryear,11pt]{elsarticle}
 \usepackage{graphics}
\usepackage{amssymb}
\usepackage{amsthm}
\usepackage{amsmath}

\usepackage[paperwidth= 21cm, paperheight= 29.7cm,
            body={15true cm,22.5true cm},
            twosideshift=2 pt,
            headheight=1.0true cm
            ]{geometry}

\begin{document}
\newtheorem{remark}{\textbf{Remark}}
\newtheorem{theorem}{\textbf{Theorem}}
\newtheorem{definition}{\textbf{Definition}}
\newtheorem{proposition}{\textbf{Proposition}}
\newtheorem{corollary}{\textbf{Corollary}}

\begin{frontmatter}



\title{Hedging strategies with a put option and their failure rates}

%\fntext[label3]{Department of Statistics and Actuarial Science, Chongqing
%University, Chongqing, 400030  P.R. CHINA;}


\author{Guanghui Huang\corref{cor1}}
\ead{hgh@cqu.edu.cn}

\author{Jing Xu}

\author{Wenting Xing}

\cortext[cor1]{Corresponding author.}
\address{College of Mathematics and Statistics, Chongqing University,
  Chongqing 401331, China}

\begin{abstract}

The problem of stock hedging is reconsidered in this paper, where a put option is chosen from a set of available put options to hedge the market risk of a stock. A formula is proposed to determine the probability that the potential loss exceeds a predetermined level of Value-at-Risk, which is used to find the optimal strike price and optimal hedge ratio. The assumptions that the chosen put option finishes in-the-money and the constraint of hedging budget is binding are relaxed in this paper. A hypothesis test is proposed to determine whether the failure rate of hedging strategy is greater than the predetermined level of risk. The performances of the proposed method and the method with those two assumptions are compared through simulations. The results of simulated investigations indicate that the proposed method is much more prudent than the method with those two assumptions.

\end{abstract}


\end{frontmatter}

% \linenumbers


\section{Introduction}

Risk management is important in the practices of financial institutions and other corporations \citep{mian1996,stulz1996,bondar1998,ahn1999,prevost2000}. Derivatives are popular instruments to hedge exposures due to currency, interest rate and other market risks \citep{berkman1997,graham2002,bartram2009,deelstra2010}. An important step of risk management is to use these derivatives in an optimal way.

The most popular derivatives are forwards, options and swaps. They are basic blocks for all sorts of other more complicated derivatives, and should be used prudently. Several parameters need to be determined in the processes of risk management, and it is necessary to investigate the influence of these parameters on the aims of the hedging policies and the possibility of achieving these goals \citep{annaert2007}.
However, the literature on risk management is much silent on how to optimally decide on these parameters \citep{annaert2007}.

The problem of determining the optimal strike price and optimal hedging ratio is considered by  \cite{ahn1999}, \cite{annaert2007}, \cite{deelstra2010} and the references therein, where a put option is used to hedge market risk under a constraint of budget. The chosen option is supposed to finish in-the-money at maturity in the aforementioned papers, such that the predicted loss of the hedged portfolio is different from the realized loss. And the constraint of hedging budget is supposed to be binding in those papers, which means that the company will always spend the maximum available to buy options, such that the cost of hedging is determined by the amount of hedging budget. Whether the performance of hedging strategy is affected by the two assumptions is considered in this paper.

Following \cite{ahn1999}, \cite{annaert2007} and \cite{deelstra2010}, the aim of hedging is to minimize the potential loss of investment under a specified level of confidence. In other words, the optimal hedging strategy is to minimize the Value-at-Risk (VaR) under a specified level of risk. However, the present paper is different from the aforementioned papers in several aspects. First, the chosen put option is not supposed to finish in-the-money at maturity in this paper. As the possibility of inexecution is taken into account, the predicted loss of hedging is closer to the realized loss. Second, the constraint of hedging budget is not supposed to be binding, such that the expenditure of hedging is not always equal to the maximum available. Third, the available put options are specified by their strike prices in a discrete manner, such that the optimal strike price can only be chosen from a predetermined finite set of strike prices, which is similar to the situation faced in real world financial market. Finally, the performances of the resulted optimal hedging strategies are investigated through hypothesis tests, where the failure of hedging means that the realized loss exceeds the level of VaR predicted by the hedging strategy. The simulated investigations indicate that the proposed method is more accurate than the method deduced from the spirit of \cite{ahn1999}, \cite{annaert2007} and \cite{deelstra2010}.

The paper is organized as  follows. Section 2 describes the stock hedging problem. Section 3 presents the main theoretical analysis of loss and its risk, where the probability that the potential loss of hedging strategy exceeds a predetermined threshold is calculated under a geometric Brownian motion. Section 4 describes two methods to determine the optimal hedging strategy, one is deduced from the spirit of the aforementioned papers, and the other is proposed in the present paper.
The failure rates of hedging strategies are compared through simulations in Section 5.  Section 6 gives the conclusions and discussions.

\section{The stock hedging problem}

Analogously to \cite{ahn1999} and \cite{deelstra2010}, a stock is supposed to be bought at time zero with price $S_0$, and to be sold at time $T$ with uncertain price $S_T$. In order to hedge the market risk of the stock, the company decides to choose one of the available put options written on the same stock with maturity at time $\tau$, where $\tau$ is prior and close  to $T$, and the $n$ available put options are specified by their strike prices $K_i$ ($i=1,2,\cdots,n$). As the prices of different put options are also different, the company needs to determine an optimal hedge ratio $h$ ($0\le h \le 1$) with respect to the chosen strike price. The cost of hedging should be less than or equal to the predetermined hedging budget $C$. In other words, the company needs to determine the optimal strike price and hedging ratio under the constraint of hedging budget.

The chosen put option is supposed to finish in-the-money at maturity, and the constraint of hedging expenditure is supposed to be binding by \cite{ahn1999},  \cite{annaert2007}, \cite{deelstra2010} and the references therein. These two assumptions are relaxed in this paper, such that the possibility of inexecution is taken into account, and the cost of hedging may be less than or equal to the maximum available. The performances of the hedging strategies with or without these assumptions are compared in the following sections.

\section{Loss and its risk}

Suppose the market price of the stock is $S_0$ at time zero, the hedge ratio is $h$, the price of the put option is $P_0$, and the riskless interest rate is $r$. At time $T$, the time value of the hedging portfolio is
\begin{equation}
S_0 e^{rT}+ h P_0 e^{rT},
\end{equation}
and the market price of the portfolio is
\begin{equation}
S_T + h\left(K-S_{\tau}\right)^{+}e^{r(T-\tau)},
\end{equation}
therefore the loss of the portfolio is
\begin{equation}
L=\left( S_0 e^{rT}+ h P_0 e^{rT} \right) - \left( S_T + h\left(K-S_{\tau}\right)^{+}e^{r(T-\tau)} \right),
\end{equation}
where $x^+=max(x,0)$, which is the payoff function of put option at maturity.

For a given threshold $v$, the probability that the amount of loss exceeds $v$ is denoted as
\begin{equation}
\alpha= \mathrm{Prob}\left\{ L \ge v \right\},
\end{equation}
in other words, $v$ is the Value-at-Risk (VaR) at $\alpha$ percentage level. There are several alternative measures of risk, such as CVaR (Conditional Value-at-Risk), ESF (Expected Shortfall), CTE (Conditional Tail Expectation), and other coherent risk measures \citep{artzner1999,rockafellar2002,acerbi2002,annaert2007,brazauskas2008,deelstra2010}. The criterion of optimality adopted in this paper is to minimize the VaR of the hedging strategy, which follows the papers by \cite{ahn1999}, \cite{annaert2007} and the references therein.

The mathematical model of stock price is chosen to be a geometric Brownian motion in this paper, following \cite{ahn1999}, i.e.
\begin{equation}
\frac{\mathrm{d}S_t}{S_t}=\mu \mathrm{dt} + \sigma \mathrm{dBt},
\end{equation}
where $S_t$ is the stock price at time $t$ ($0< t \le T$), $\mu$ and $\sigma$ are the drift and the volatility of stock price, and $B_t$ is a standard Brownian motion. The solution of the stochastic differential equation is
\begin{equation}
S_t = S_0 e^{\sigma B_t + \left( \mu-\frac{1}{2}\sigma^2 \right)t},
\end{equation}
where $B_0 = 0$, and $S_t$ is lognormally distributed.

\begin{proposition}\label{proposition1}
For a given threshold of loss $v$, the probability that the loss exceeds $v$ is
\begin{equation}
\mathrm{Prob}\left\{ L \ge v \right\} = {E}\left[ I_{\left\{X \le c_1 \right\}}  F_{Y} \left( g(X)-X \right) \right]
  +
  {E}\left[ I_{\left\{X \ge c_1 \right\}} F_{Y} \left( c_2-X \right) \right],
\end{equation}
where $E[X]$ is the expectation of random variable $X$. $I_{\left\{X<c\right\}}$ is the index function of $X$ such that $I_{\left\{X<c\right\}}=1$ when $\{X<c\}$ is true, otherwise  $I_{\left\{X<c\right\}}=0$. $F_{Y}(y)$ is the cumulative distribution function of random variable $Y$, and
\begin{eqnarray}
c_1 & = &
   \frac{1}{\sigma}\left[ \ln\left(\frac{K}{S_0}\right)- \left(\mu - \frac{1}{2} \sigma^2 \right) \tau \right], \nonumber \\
g(X) & = &
    \frac{1}{\sigma}
     \left[
          \ln \left(
             \frac{\left( S_0 + hP_0\right)e^{rT} - h \left( K - f(X)  \right)e^{r(T-\tau)} - v}{S_0}
          \right)
           - \left(
            \mu - \frac{1}{2} \sigma^2
            \right) T
     \right],\nonumber\\
f(X) & = &
    S_0 e^{\sigma X + \left( \mu - \frac{1}{2}\sigma^2 \right)\tau}, \nonumber \\
c_2 & = &
          \frac{1}{\sigma}
     \left[
          \ln \left(
             \frac{\left( S_0 + hP_0\right)e^{rT} - v}{S_0}
          \right)
           - \left(
            \mu - \frac{1}{2} \sigma^2
            \right) T
     \right].\nonumber
\end{eqnarray}
$X$ and $Y$ are both normally distributed, where $X \sim N(0,\sqrt{\tau})$, $Y \sim N(0, \sqrt{T-\tau})$.
\end{proposition}
Proof: see Appendix.

\begin{remark}
For a specified hedging strategy, $Q(v)=\mathrm{Prob}\left\{ L \ge v \right\}$ is a decreasing function of $v$. The $VaR$ under $\alpha$ level can be obtained from equation
\begin{equation} \label{qequation}
Q(v)=\alpha.
\end{equation}
The expectations in Proposition \ref{proposition1} can be calculated with Monte Carlo simulation methods, and the optimal hedging strategy which has the smallest VaR can be obtained from equation (\ref{qequation}) by numerical searching methods.
\end{remark}

\section{Optimal hedging strategies}

Suppose there are $n$ put options available in the market with different strike prices $K_i$, $i=1,2,\cdots,n$, and the prices of those put options are denoted as $P_i$ respectively, which are determined by the Black-Scholes formula
\begin{equation}
P_i=K_i e^{-r\tau}N(d_1)-S_0 N(d_2),
\end{equation}
where
\begin{equation}
d_1=\frac{\ln\left(\frac{K_i}{S_0}\right)-\left(r-\frac{1}{2}{\sigma}^2\right)\tau}
   {\sigma\sqrt{\tau}},
\end{equation}
\begin{equation}
d_2=\frac{\ln(\frac{K_i}{S_0})-\left(r+\frac{1}{2}{\sigma}^2\right)\tau}
   {\sigma\sqrt{\tau}}=d_1 -\sigma\sqrt{\tau},
\end{equation}
and $N(\cdot)$ is the cumulative distribution function of standard normal distribution.

In order to hedge market risk,
the company decides to buy one or part of a put option with an optimal strike price, therefore how to determine the optimal strike price and optimal hedging ratio is important in the practice of risk management. The present paper investigates the performances of two kinds of hedging strategies which are based on different assumptions.

\subsection{ABRW method}

A method to determine the optimal strike price and optimal hedge ratio is given by
\cite{ahn1999}, \cite{annaert2007} and \cite{deelstra2010},  where the chosen put option is supposed to finish in-the-money at maturity, and the company will always spend the maximum available to buy put option,  i.e.
\begin{equation}
\min_{\left(h,K_i\right)} VaR_{\alpha} = \min_{(h,K_i)}  \left(S_0+h P_{i}\right) e^{rT} -
\left[
 (1-h)S_0 e^{(\mu-\frac{1}{2}\sigma^2)T+\theta(\alpha)\sigma\sqrt{T}}+h K_i e^{r(T-\tau)}\right],
\end{equation}
\begin{eqnarray}
\text{subject   to }\left\{
\begin{array}{l} h P_{i}=C,   \\
P_i=K_i e^{-r\tau}N(d_1)-S_0 N(d_2),\\
  0\leq h\leq 1,   \\
   K_i \in \left\{K_1, K_2, \cdots, K_n\right\},
   \end{array}
   \right.
\end{eqnarray}
where stock price is supposed to be a geometric  Brownian motion,  $C$ is the maximum amount available to buy put option, and $\theta(\alpha)$ is the cut-off point of the cumulative distribution function of standard normal distribution. The solution of this problem is achieved through numerical searching method in this paper, which is denoted as $\left(K^*,h^*\right)$, and called the optimal strategy of ABRW method.

\subsection{Minimization of VaR}

Following \cite{ahn1999}, the subject of hedging is to minimize the VaR with one of the available put options. When the inexecution of the put option is taken into account, and  the expenditure of hedging is not always the same as the maximum available,  the probability that the potential loss exceeds the specified value of VaR can be calculated by Proposition \ref{proposition1}. In other words, the company wants to solve the following problem
\begin{eqnarray}
\min_{\left(h,K_i\right)}VaR_{\alpha},\\
\text{subject  to}
\left\{
\begin{array}{l}
\mathrm{Prob}\left\{ L \ge VaR_{\alpha} \right\}=\alpha,   \\
hP_{i} \le C,   \\
P_i=K_i e^{-r\tau}N(d_1)-S_0 N(d_2),\\
0 \le h \le 1,    \\
K_i \in \left\{K_1, K_2, \cdots, K_n \right\},
\end{array}
\right.
\end{eqnarray}
where $\alpha$ is the specified target of risk control, and $C$ is the constraint of hedging budget. The solution of this problem is denoted as $\left( \overline{K}^*, \overline{h}^* \right)$, which is called the optimal hedging strategy with minimum VaR under $\alpha$ percentage level.  Numerical searching method is used to solve this optimal problem in this paper.

\section{Simulated investigations}

For the same level of $\alpha$,
the resulted VaR deduced by ABRW method may be different from the one proposed in this paper. A hedging strategy is called success if the realized loss is less than or equal to the predetermined level of VaR, otherwise the hedging strategy is called failure. In order to compare the performances of ABRW method and the proposed method, simulated investigations are designed in this paper.

The failure rates under various parameters are considered in this paper. The parameters are divided into four parts, including market factors $\mu$ and $\sigma$ (/year), time factors $\tau$ and $T$ (day), management factors $C$ and $\alpha$, and interest rate factor $r$ (/year). The standard value of each parameter is the fixed value when the other parameters are varied. The standard value, range of variation and step length of each parameter are given in Table \ref{parameters}, and the stock price at time zero $S_0$ is fixed at $100$ in this paper.

\begin{table}
\begin{center}
\caption{Standard value, range of variation and step length of each parameter.}
\label{parameters}
\begin{tabular}{lcccccccc}
\hline
& $\mu$  & $\sigma^2$ &      $\tau$  & $T-\tau$ &      C      &     $\alpha$  &  r\\
\hline
standard &  0.1         &     0.0225         &       35      &        5               &     0.35    & 0.05 &  0.05 \\

range    &[-0.1,0.1]    & [0.001,0.21]     &    [1,40]     &     [1,40]             &  [0.05,5]   & [0.01,0.05]   & [0.01,0.16]\\
step   &  0.01   &     0.001        &     1         &       1                &    0.05     &    0.005      & 0.001\\
\hline
\end{tabular}
\end{center}
\end{table}

\subsection{Test of failure ratio}

The frequency of failure is used to evaluate the performance of hedging strategy. For a particular parameter combination, the optimal strike price, optimal hedging ratio and minimum VaR under $\alpha$ level  are determined by ABRW method and the proposed method respectively. The stock prices $S_{\tau}$ and $S_{T}$ are simulated with geometric Brownian motion, and the hedging is said to be failure when the realized loss exceeds the predetermined level of VaR.
Repeat the simulated trial $N=100,000$ times, the probability of failure can be estimated by the observed frequency of failures. If the probability of failure is less than or equal to the specified value of $\alpha$, the hedging strategy is said to be success, otherwise it is said to be failure.

A hypothesis test is used to determine whether the probability of failure is larger than the specified value of $\alpha$. Denote the probability of failure as $f$, the null and alternative hypotheses are
\begin{equation}
H_0: f \le \alpha; \quad H_1: f > \alpha.
\end{equation}

Let $X_i$ be the index variable of the $i$th trial, such that $X_i=1$ indicates that the hedging is failure at the $i$th trial, otherwise $X_i=0$, $i=1,2,\cdots,N$, where $N$ is the number of simulations. The frequency of failures is
\begin{equation}
\overline{X}=\frac{1}{N} \sum_{i=1}^{N} X_i.
\end{equation}
The test statistic is chosen to be
\begin{equation}
T=\frac{\overline{X}-\alpha}{\sqrt{\alpha\left(1-\alpha\right)/N}},
\end{equation}
which is asymptotically normally distributed, therefore
\begin{equation}
K = \left\{ T \ge U_{1-\beta} \right\}
\end{equation}
is used as the rejection criterion for the test, where $U_{1-\beta}$ is the ${1-\beta}$ quantile of cumulative distribution function of standard normal distribution, and $\beta=0.05$ is the significance level of the test in this paper.

From Table \ref{passingrate}, it can be found that the passing rates of the hedging strategy deduced by ABRW method are dramatically lower than their counterparts deduced by the proposed method.
These observations indicate that the hedging strategy given by the proposed method is much more prudent than the strategy given by the ABRW method with respect to the failure rates.

\begin{table}
\begin{center}
\caption{Passing rates of hypothesis tests under various parameter combinations.}
\label{passingrate}
\begin{tabular}{ccccccccccccc}
\hline
 & $\mu$ & $\sigma$  &    &  $\tau$ &  $T - \tau$ &    &  C & $\alpha$ &   &  r  &  & Total\\
 \cline{2-3} \cline{5-6} \cline{8-9} \cline{11-11} \cline{13-13}
number of combinations      &  \multicolumn{2}{c}{4410} &  & \multicolumn{2}{c}{1600} & &  \multicolumn{2}{c}{900}   &     &  1501 & & 8409\\
\hline
ABRW               &  \multicolumn{2}{c}{0.84\%} & & \multicolumn{2}{c}{2.13\%} & &  \multicolumn{2}{c}{0.56\%}  &  &  0\% & & 0.9\%
\\
Min VaR          & \multicolumn{2}{c}{100\%} &  & \multicolumn{2}{c}{100\%} & &  \multicolumn{2}{c}{100\%} &    & 100\% & & 100\%
\\
\hline
\end{tabular}
\end{center}
\end{table}

\subsection{Failure rate and hedging budget}

\begin{figure}
  \centering
  \includegraphics[width=0.8\textwidth]{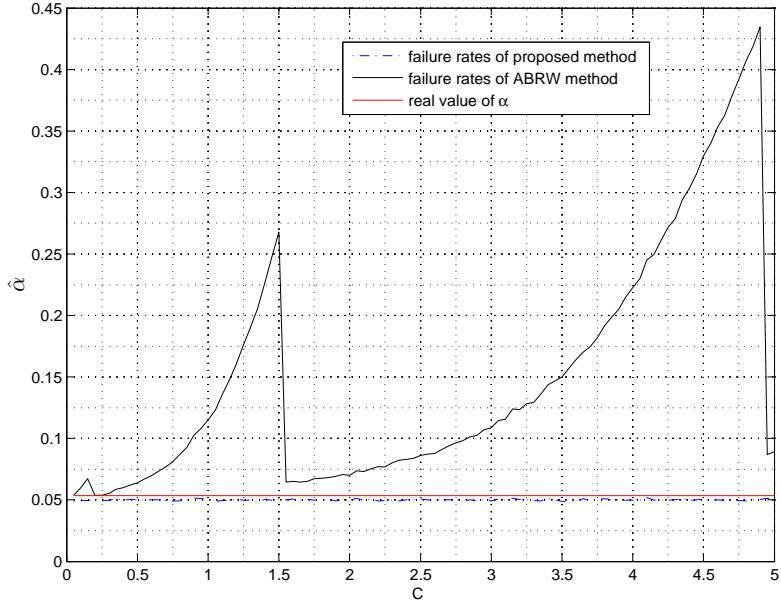}
  \caption{Failure rates and hedging budget.}\label{figure}
\end{figure}

The constraint on hedging expenditure is supposed to be binding by \cite{ahn1999}, \cite{annaert2007} and \cite{deelstra2010}, in other words, the company will always spend the maximum available to buy put option. In order to investigate the effect of this assumption,
the probabilities of failure are estimated from the simulated data, where $C$ increases from 0.05 to 5, and the results are plotted in Figure \ref{figure}.

It can be found that the estimated probabilities of failure under the proposed method are all less than the predetermined level of risk management, while their counterparts under the ABRW method are all greater than the specified level of $\alpha$.
Furthermore, the probability of failure with lower level of $C$ is not always greater than the probability with higher level of hedging budget under ABRW method. These observations indicate that more hedging cost may not always produce better consequence of risk management.

\subsection{Numerical examples}

Let all of the parameters take the standard values, the optimal strike prices and hedging ratios under both hedging strategies are the same, where $K=100$ and $h=0.231$. The VaR under ABRW method is $5.8636$, and the failure rate is $0.0563$, which does not pass the hypothesis test. And the VaR under the proposed method is $6.1223$, and the failure rate is $0.04825$, which passes the test.

If the influence of the binding constraint on hedge budget of ABRW method is absolutely eliminated, the optimal strike price is $K=95$, and the optimal hedging ratio is $h=1$. The failure ratio reaches to a higher level $0.07595$. Therefore the worse result is just derived from the assumption that the put option always finishes in-the-money.

Let $K=105$, $h=0.071293$, the corresponding VaR is $6.5666$ under the ABRW method,  and the rate of failure is $0.05025$. Although this hedging strategy is not optimal in the sense of ABRW, the performance is better than the optimal one given by ABRW method. This observation indicates that the two key assumptions taken by \cite{ahn1999}, \cite{annaert2007} and \cite{deelstra2010} may lead to  strategies  which is not really optimal with respect to the test of failure rate.

\section{Conclusions and discussions}

The stock hedging problem is considered in this paper, where a put option is chosen from a finite set of available options to manage the market risk exposure of a stock. A formula is proposed  to determine the probability that the potential loss exceeds a predetermined level of VaR under geometric Brownian motion, where market risk is measured with Value-at-Risk, and the assumptions that the chosen put option finishes in-the-money and the constraint of hedging budget is binding are relaxed.
A method to determine the optimal strike price and optimal hedging ratio is proposed in this paper.

The performances of the proposed method and the method deduced from the spirit of \cite{ahn1999}, \cite{annaert2007} and \cite{deelstra2010} are compared through simulated investigations, where a hypothesis test is proposed to determine whether the failure rate of hedging strategy is greater than the specified level of risk.
The results of simulations indicate that the proposed method is much more prudent  than the method deduced from the aforementioned papers.

The differences between the two methods are derived from the two assumptions. The chosen put option is supposed to be in-the-money at maturity,  such that the predicted loss of hedging is different from the realized loss in financial market. And the constraint of budget is supposed to be binding in the aforementioned papers,  which means that the company will always spend the maximum available to hedge market risk, such that the cost of hedging is affected by the predetermined amount of budget. The tests of failure rates indicate that the performances of the hedging strategy without the two assumptions are much more prudent than the strategy deduced from those assumptions.

Further research  possibilities are mainly in three directions. First, other model of stock price can be used to determine the potential loss of hedging strategy. The use of geometric Brownian motion to describe the dynamics of stock prices is very widespread in the financial industry, while empirical investigations indicate that the returns of stock prices are not always lognormally distributed. Second, other models of risk measurement can be used to determine the risk of hedging strategy. VaR is not a coherent measure, while CVaR, ESF, CTE, and other coherent measures can be used as the risk measurement. Finally, the performance of the proposed method in real word financial market is also needed to be investigated, which will be left for further research.


\appendix
\section*{Appendix}
\begin{proof}
The probability that the potential loss exceeds $v$ can be written as
\begin{align*}
& \mathrm{Prob}\left\{ S_0e^{rT}+hP_0e^{rT}-h(K-S_{\tau})^+e^{r(T-{\tau})}-S_T\ge v \right\}\\
        & =
  \mathrm{Prob}\left\{
    S_0e^{rT}+hP_0e^{rT}-h(K-S_{\tau})^+e^{r(T-{\tau})}-S_T \ge v, K>S_{\tau}
      \right\}\\
        &
        +
        \mathrm{Prob}\left\{
         S_0e^{rT}+hP_0e^{rT}-h(K-S_{\tau})^+e^{r(T-{\tau})} -S_T \ge v, K \le S_{\tau}
        \right\}\\
        &
        \triangleq Q_1+Q_2.
\end{align*}
Set $X=B_{\tau}$, $Y=B_T-B_{\tau}$. As $B_{t}$ $ \left( 0< t \le T \right)$ is a standard Brownian  motion, $X$ and $Y$ are independently distributed, and $X \sim N(0,\sqrt{\tau}),Y\sim N(0,\sqrt{T-{\tau}})$.

\begin{align*}
Q_1
    &
    =
    \mathrm{Prob}\left\{
     S_T \le  S_0e^{rT}+hP_0e^{rT}-h(K-S_{\tau})e^{r(T-{\tau})}-v, S_{\tau} < K
     \right\}\\
    &=
     \mathrm{Prob}\left\{
     B_T \le
      \frac{1}{\sigma}
      \left(
       \ln\left( \frac{\left( S_0 + h P_0\right)e^{rT}- h \left(K-S_{\tau}\right) e^{r \left( T-\tau \right)}- v}{S_0}\right)-\left(\mu - \frac{1}{2}\sigma^2 \right)T
      \right), \right. \\
      &
       \left.
     \quad B_{\tau} \le  \frac{1}{\sigma} \left( \ln\left(\frac{K}{S_0}\right) - \left(\mu - \frac{1}{2}\sigma^2\right)\tau \right)
     \right\}\\
    & =\mathrm{E}\left[ I_{\left\{X \le c_1 \right\}}  F_{Y} \left( g(X)-X \right) \right],
\end{align*}
where $F_{Y}(y)$ is the distribution function of variable $Y$, and
\begin{eqnarray}
c_1 & = &
   \frac{1}{\sigma}\left[ \ln\left(\frac{K}{S_0}\right)- \left(\mu - \frac{1}{2} \sigma^2 \right) \tau \right], \nonumber \\
g(X) & = &
    \frac{1}{\sigma}
     \left[
          \ln \left(
             \frac{\left( S_0 + hP_0\right)e^{rT} - h \left( K -f(X)  \right)e^{r(T-\tau)} - v}{S_0}
          \right)
           - \left(
            \mu - \frac{1}{2} \sigma^2
            \right) T
     \right],\nonumber\\
f(X) & = &
    S_0 e^{\sigma X + \left( \mu - \frac{1}{2}\sigma^2 \right)\tau}. \nonumber
\end{eqnarray}

\begin{align*}
Q_2
    & =\mathrm{Prob} \left\{
      S_T \le  S_0e^{rT} + hP_0 e^{rT} - v,  S_{\tau} \ge  K
      \right\} \\
    & = \mathrm{Prob}\left\{
     B_T \le \frac{1}{\sigma} \left( \ln \left( \frac{\left(S_0 + hP_0\right)e^{rT}-v}{S_0} \right) - \left(\mu-\frac{1}{2}\sigma^2\right)T  \right), \right.\\
     & \quad
     \left.
      B_{\tau} \ge   \frac{1}{\sigma} \left( \ln\left(\frac{K}{S_0}\right) - \left( \mu - \frac{1}{2}\sigma^2\right) \tau \right)
     \right\}\\
    &={E}\left[ I_{\left\{X \ge c_1 \right\}} F_{Y} \left( c_2-X \right) \right],
\end{align*}
where
\begin{eqnarray}
c_2 & = &
          \frac{1}{\sigma}
     \left[
          \ln \left(
             \frac{\left( S_0 + hP_0\right)e^{rT} - v}{S_0}
          \right)
           - \left(
            \mu - \frac{1}{2} \sigma^2
            \right) T
     \right].\nonumber
\end{eqnarray}
\end{proof}

% Acknowledgments here
\section*{Acknowledgments}
 {Guanghui Huang is supported by the Fundamental Research Funds for the Central Universities of China, Project No. CDJZR10 100 007.}

\bibliographystyle{elsarticle-harv}



\end{document}